\documentclass[12pt]{article}

\usepackage{latexsym,amsfonts,amsmath,amsthm,amssymb}

\begin{document}




\title{On $p$-Adic Mathematical Physics }

\author{B.~Dragovich$^1$, A.~Yu.~Khrennikov$^2$, S.~V.~Kozyrev$^3$ and
I.~V.~Volovich$^3$ \\ {}\\ $^1$Institute of Physics, Pregrevica
118, 11080 Belgrade, Serbia \\  $^2$V$\ddot{a}$xj$\ddot{o}$
University, V$\ddot{a}$xj$\ddot{o}$, Sweden \\  $^3$Steklov
Mathematical Institute, ul. Gubkina 8, Moscow, 119991 Russia }



\date{~}

\maketitle

\begin{abstract}
A brief review of some selected topics in $p$-adic mathematical
physics is presented.
\end{abstract}














\section{{Numbers: Rational, Real, $p$-Adic}}

We present a brief review of some selected topics in $p$-adic
mathematical physics. More details can be found in the references
below and the other references are mainly contained therein. We hope
that this brief introduction to some aspects of $p$-adic
mathematical physics could be helpful for the readers of the first
issue of the journal {\it $p$-Adic Numbers, Ultrametric Analysis and
Applications}.

The notion of numbers is basic not only in mathematics but also in
physics and entire science. Most of modern science is based on
mathematical analysis over real and complex numbers. However, it
is turned out that for exploring complex hierarchical systems it
is sometimes more fruitful to use analysis over $p$-adic numbers
and ultrametric spaces. $p$-Adic numbers (see, e.g. \cite{Ser}),
introduced by Hensel, are widely used in mathematics: in number
theory, algebraic geometry, representation theory, algebraic and
arithmetical dynamics, and cryptography.

The following view  how to do science with numbers has been put
forward by Volovich in \cite{Vol1,Volovich:1987}. Suppose we have a
physical or any other system and we make measurements. To describe
results of the measurements, we can always use rationals. To do
mathematical analysis, one needs a completion of the field
$\mathbb{Q}$ of the rational numbers. According to the Ostrowski
theorem there are only two kinds of completions of the rationals.
They give real $\mathbb{R}$ or $p$-adic  $\mathbb{Q}_p$ number
fields, where $p$ is any prime number with corresponding $p$-adic
norm $|x|_p$, which is non-Archimedean. There is an adelic formula
$\prod_p|x|_p=1$ valid for rational $x$ which expresses the real
norm $|x|_{\infty}$ in terms of $p$-adic ones. Any $p$-adic number
$x$ can be represented as a series $x=\sum_{i=k}^{\infty}a_ip^i$,
where $k$ is an integer and $a_i \in \{0,1,2,...,p-1\}$ are digits.
To build a mathematical model of the system we  use real or $p$-adic
numbers or both, depending on the properties of the system
\cite{Vol1,Volovich:1987}.

Superanalysis over real and $p$-adic numbers has been considered by
Vladimirov and Volovich \cite{VL,VL1}. An adelic approach was
emphasized by Manin \cite{Manin}.

One can argue that at the very small (Planck) scale the geometry of
the spacetime should be non-Archimedean
\cite{Vol1,Volovich:1987,Var}. There should be quantum fluctuations
not only of metrics and geometry but even of the number field.
Therefore, it was suggested \cite{Vol1} the following {\it {number
field invariance principle}}: Fundamental physical laws should be
invariant under the change of the number field.

One could start from the ring of integers or the Grothendieck
schemes. Then rational, real or $p$-adic numbers should appear
through a mechanism of number field symmetry breaking, similar to
the Higgs mechanism \cite{bogol,nambu}.

Recently (for a review, see
\cite{VVZ,Khrennikov:1994,BrFr,Koch,KozyrevBook}) there have been
exciting achievements exploring $p$-adic, adelic and ultrametric
structures in various models of physics: from the spacetime geometry
at small scale and strings, via spin glasses and other complex
systems, to the universe as a whole. There has been also significant
progress in non-Archimedean modeling of some biological, cognitive,
information and stochastic phenomena.

Ultrametricity seems to be a generic property of complex systems
which contain hierarchy. Moreover, there is some evidence towards
much more wide applicability of $p$-adic and non-Archimedean methods
to various fields of knowledge. To extend $p$-adic methods into
actual problems in diverse fields of economics, medicine,
psychology, sociology, control theory, as well as to many other
branches of sciences, is a great challenge and a great opportunity.

\section{{$p$-Adic Strings}}

String theory is a modern  unified theory of elementary particles
\cite{GSW}.

$p$-Adic string theories with $p$-adic valued and also with complex
valued amplitudes were suggested by Volovich in
\cite{Vol1,Volovich:1987}.  These two possibilities are in
correspondence with two forms of the dual string amplitude mentioned
in \cite{Vol1}. The first one uses the Veneziano amplitude, which
describes scattering of elementary particles, in the form
$$
A (a,b) = \frac{\Gamma (a) \,  \Gamma (b)}{\Gamma (a+b)} ,
$$
where $a$ and $b$ are parameters depending on momenta of colliding
particles. As a $p$-adic Veneziano amplitude it was suggested
$$
A_p (a,b) = \frac{\Gamma_p (a) \,  \Gamma_p (b)}{\Gamma_p (a+b)} ,
$$
where $\Gamma_p$ is the $p$-adic valued  Morita gamma function. The
second one uses the following form of the Veneziano amplitude
$$
A (a,b) = \int_0^1  x^{a-1}\, (1 - x)^{b-1}\, dx .
$$
 Since the function $x^a$ is a multiplicative character on the real
axis we can interpret the Veneziano amplitude as {\it the
convolution of two characters}.  In \cite{Vol1} an analogue of the
crossing symmetric Veneziano amplitude on the Galois field $F_p$ is
also introduced as the convolution of the corresponding
complex-valued characters, which is the Jacobi sum,
$$
A(a,b)=\sum_{x\in F_p}\chi_a (x)\, \chi_b (1-x).
$$

Freund and Olson  \cite{FrOl} introduced an analogue of the crossing
symmetric Veneziano amplitude on the $\mathbb{Q}_p$  as the
convolution of the corresponding complex-valued characters, which is
the Gel'fand-Graev beta  function \cite{GGP},
$$
A(a,b)=\int_{\mathbb{Q}_p}\chi_a (x)\, \chi_b (1-x)\, dx .
$$

Frampton and Okada
 \cite{fram1}, and Brekke, Freund, Olson and Witten \cite{freund} have discovered that  string amplitudes
 given by the above formula and its generalization can be described by  nonlocal effective field theories.
 Important
adelic formulas were considered by Freund and Witten \cite{FroWit},
see also \cite{VolHar}. Vladimirov found a unified approach to
adelic formulas for the superstring amplitudes using algebraic
extensions \cite{borevich} of the number fields \cite{Vlaade} (see
also \cite{ADV1,ADV2,ADV3}).

Loop corrections to the $p$-adic string amplitudes are considered in
\cite{leb,chek}

More information on $p$-adic string theory see in
\cite{FraOka,nishino,BrFr,VVZ,volovich1}.

Strings, motives and $L$-functions are discussed by Volovich
\cite{Volmot}. String partition function can be expressed as inverse
to the Mellin transform of $L$-function of the Deligne motive,
$$
L(s)=\sum_n\tau (n)n^{-s},
$$
where $\tau (n)$ is the Ramanujan function.

Motives and quantum fields are discussed by Connes and Marcolli
\cite{marcolli1}. Motives, algebraic and noncommutative geometry are
explored in
\cite{marcolli2,marcolli3,marcolli4,marcolli5,marcolli6}. Theory of
motives is considered by Voevodsky \cite{voevodsky}.

$p$-Adic geometry is dicsussed in \cite{thakur}.

\section{{$p$-Adic Field Theories}}

As a free action in $p$-adic field theory one can take the following
functional
$$
 S(f)=\int_{\mathbb{Q}_p}f D f dx
 $$
 where $f = f (x)$ is a function $f : \mathbb{Q}_p \to \mathbb{R}$, $dx$ is the Haar measure
 and $D$ is the Vladimirov operator or its generalizations \cite{Koch}.
A $p$-adic analog of the Euclidean quantum field theory was
introduced by Kochubei and Sait-Ametov \cite{[A]}.

 Renormalizations in $p$-adic field theory are studied by Missarov \cite{Mis} and  Smirnov
 \cite{Smi}.

Nonlocal scalar field theory \cite{fram1,freund} for $p$-adic
strings has occurred to be very instructive toy model
\cite{sen,zwiebach} for truncated string field theory models
\cite{ABGKM}. Boundary value problems for homogeneous solutions of
  nonlinear equations of motion corresponding to the  $p$-adic string \cite{fram1,freund},
  $$
  e^{\Box}\Phi=\Phi^{p}
  $$
  and its generalizations were explored in
  \cite{Yaroslav,VSV-Yaroslav,AJK,Vlanon,VlaFirst}   (here $\Box$ is the
d'Alembert operator, the field $\Phi$ and its argument are real-valued).
  A
  truncated version of the supersting
  field theory describing non-BPS branes \cite{ABKM}
  $$
  (-\Box +1)e^{\Box}\Phi=\Phi^{3}
  $$
  was investigated in
  \cite{Yaroslav,AJK,Proxor,Trento}.

  A generalization of these nonlocal models
  to the case of curved spacetime has been proposed by Aref'eva \cite{Are-cosmology}
  and their  application to cosmology, in particular,
  to the inflation and
dark energy  was initiated. Cosmological applications have been
studied in numerous  papers
\cite{Are-Jouk,AK,Calcagni,dragov13,Cline,Koshel,Jouk-PR,CMN,Are-Jouk-Ver,Neil,Mulryne,Neil2}.

Quantization of the Riemann zeta-function and applications to
cosmology are considered in \cite{AreVol}. If $\zeta(s)$ is the
Riemann zeta-function then the quantum zeta function is the
pseudodifferential operator $\zeta(\Box)$, i.e. the Riemann
zeta-function is the symbol of the pseudodifferential operator
$\zeta{(\Box})$. Quantization of the Langlands program is also indicated.

There is a recent consideration towards an effective Lagrangian for
adelic strings with the Riemann zeta function nonlocality
\cite{dragov10}.

Zeta-functions and many other areas of mathematics are considered
by Shai Haran in \cite{haran}.

A dual connection between $p$-adic analysis and noncommutative geometry
was pointed out   by Aref'eva and Volovich \cite{AreVolNC}.
 It was found that the Haar measure on quantum group $SU_q(2)$ is
equivalent to the Haar measure on $p$-adic line $\mathbb{Q}_p$ if $q=1/p$.

Relation between the space of coherent states for free annihilation
operators and the space of $p$--adic distributions was found in \cite{kozy}.

Physics, number theory,  and noncommutative geometry are discussed
by Connes and Marcolli \cite{marcolli3a,marcolli4a}.

The number field invariance principle \cite{Vol1} requires
consideration of quantum  field theory on an arbitrary number field.
Such consideration is attempted in \cite{schmidt}.

\section{{$p$-Adic and Adelic Quantum Mechanics}}

A general modern approach to quantum theory is presented in the
Varadarajan book \cite{raja1}. There are several versions of
$p$-adic quantum mechanics. One formulation with the complex
valued wave functions given by Vladimirov, Volovich and Zelenov
\cite{VVZ2,VVZ3} is based on the triple $\{L^2(\mathbb{Q}_p),
W(z), U(t)\}$, where $W(z)$ is a unitary representation  of the
Heisenberg-Weyl group in the Hilbert space $L^2(\mathbb{Q}_p)$ and
$U(t)$  is a unitary dynamics. Representation of canonical
commutation relations for $p$-adic quantum mechanics for finite
and infinite dimensional systems were studied by Zelenov
\cite{zelen1}. Representation theory of $p$-adic groups is
discussed in \cite{GGP,harish-chandra,weil,schneider}.

An approach to a unified $p$-adic and real theory of commutation
relations is developed by Zelenov \cite{zelenFirst}. It is based on
the interpretation of the group of functions with values in the
field of rational numbers as the experiment data space.

The consequences for particle classification of the hypothesis that
spacetime geometry is non-archimedean at the Planck scale are
explored by Varadarajan \cite{varadFirst}. The multiplier groups and
universal topological central extensions of the $p$-adic Poincar\'e
and Galilean groups are determined.

$p$-Adic Maslov index was constructed  by Zelenov \cite{zelmas}.

Another formulation \cite{VV1} uses pseudodifferential operators and
spectral theory. The free wave function of $p$-adic quantum
mechanics satisfies a pseudodifferential equation of Schr\"odinger
type. A theory of the Cauchy problem for this equation is developed
by Kochubei \cite{[B],Koch} and Zuniga-Galindo \cite{[F]}.
 Matrix valued Schr\"odinger operator on local fields
is considered in \cite{digernes}.

Adelic quantum mechanics, which is a generalization of $p$-adic and
ordinary quantum mechanics as well as their unification,  was
introduced by Dragovich \cite{dragov1}. It was found that adelic
harmonic oscillator is related to the Riemann zeta function
\cite{dragov2}. Many adelic physical systems have been considered
(as a review, see \cite{dragov3}). As a result of adelic approach
and  $p$-adic effects, there is some discreteness of space and time
in adelic systems \cite{dragov4}. Distributions on adeles are
considered by Dragovich \cite{dragov17}, E. M. Radyna and Ya. V.
Radyno \cite{radyno}.

$p$-Adic path (functional) integrals are considered by Parisi
\cite{Parisi}, Zelenov \cite{zelen2}, Varadarajan
\cite{varadarajan}, Smolyanov and Shamarov \cite{smolyanov}.
Dragovich \cite{dragov1,dragov7} introduced adelic path integral and
elaborated it with Djordjevi\'c and Ne\v si\'c \cite{dragov8}. Using
path integral approach, the probability amplitude for
one-dimensional $p$-adic quantum-mechanical systems with quadratic
Lagrangians was calculated in the exact form, which is the same as
that one in ordinary quantum mechanics \cite{dragov9}. $p$-Adic Airy
integrals are considered in \cite{fernandez}.

$p$-Adic quantum mechanics with $p$-adic valued wave functions is
reviewed below.

\section{{$p$-Adic and Adelic Gravity and Cosmology }}

$p$-Adic gravity and the wave function of the Universe are
considered by Aref'eva, Dragovich, Frampton and Volovich
\cite{ADFV}. In particular, $p$-adic Einstein equations
$$
R_{\mu \nu} - \frac{1}{2}\, R g_{\mu\nu} = \kappa\, T_{\mu \nu} -
\Lambda\, g_{\mu\nu}
$$
are explored, where $g_{\mu\nu}$ is $p$-adic valued gravitational
field. Summation on algebraic varieties and adelic products in
quantum gravity are investigated.

Adelic quantum cosmology, as an application of adelic quantum
mechanics to minisuperspace cosmological models of the very early
universe, is initiated by Dragovich \cite{dragov7,dragov5} and
developments are presented in \cite{DDNV,dragov6}. It is
illustrated by a few cosmological models, and some discreteness of
the minisuperspace and the cosmological constant are found.

As was mentioned above, nonlocal scalar field theories for
$p$-adic strings have occurred to be interesting in cosmology, in
particular, in the context of their relation with nonlocal string
field inspired models
\cite{Are-cosmology,Are-Jouk,AK,Calcagni,dragov13,Cline,Koshel,Jouk-PR,CMN,Are-Jouk-Ver,Neil,Mulryne,Neil2}.

\section{{$p$-Adic Stochastic Processes}}

The $p$-adic diffusion (heat) equation
$$
{\partial f(x,t)\over \partial t}+D^{\alpha}_x f(x,t)=0
$$
was suggested in \cite{VVZ}, and its mathematical properties and
properties of the associated stochastic processes were studied (see
also \cite{[G],[H],[I],varadarajan}). Here $f=f(x,t)$ is a real
valued function of the real time $t$ and the $p$--adic coordinate
$x$. $D^{\alpha}_x$ is the Vladimirov operator.

$p$-Adic Brownian motion is explored in
\cite{BikVol,randomfield,kamizono}, see also \cite{evans}. Various classes of
$p$-adic stochastic processes are investigated by Albeverio and
Karwowski \cite{karwowski}, Kochubei \cite{Koch}, Yasuda
\cite{yasuda1}, Albeverio and Belopolskaya \cite{belopolskaya}; see
\cite{Koch} for further references. For the most recent results on
$p$-adic stochastic integrals and stochastic differential equations
see \cite{[C]}.

Kaneko \cite{KanFirst} showed a relationship between Besov space and
potential space, and pointed out a probabilistic significance of the
relationship in terms of fractal analysis.

\section{{Vladimirov operator}}


The Vladimirov operator \cite{vladUMN,VVZ} of $p$-adic fractional
differentiation is defined as
$$
D^{\alpha} f(x)=\int_{\mathbb{Q}_p}
\chi(-kx)|k|_p^{\alpha}dk\widetilde{f}(k)dk,\qquad
\widetilde{f}(k)=\int_{\mathbb{Q}_p} \chi(kx)f(x)dx .
$$
Here $f (x)$ is a complex-valued function of $p$-adic argument $x$ and
$\chi$ is the additive character: $\chi(x)=\exp \left(2\pi i
\{x\}\right)$, $\{x\}$ is the fractional part of $x$.

For $\alpha>0$ the Vladimirov operator has the following integral
representation
$$ D^{\alpha} f(x)=\frac{1}{\Gamma_p(-\alpha)}
\int_{\mathbb{Q}_p}\frac{f(x)-f(y)}{|x-y|_p^{1+\alpha}}dy,
$$
with the constant
$$\Gamma_p(-\alpha)={p^{\alpha}-1\over 1-p^{-1-\alpha}}.$$

There is a well-developed theory of the Vladimirov operator and
related constructions (spectral properties, operators on bounded
regions, analogs of elliptic and parabolic equations, a wave-type
equation etc); see section on wavelets and \cite{VVZ,Koch,[D],[E],torba}.

   \section{{Dynamics and Evolution
   of Complex Biological
   Systems}}

An idea of using ultrametric spaces to describe the states of
complex biological systems, naturally possess a hierarchical
organization, has been sound more than once as from the middle of
the 1980th by G. Frauenfelder, G. Parisi, D. Stain, and the others, see \cite{RTV}.
In protein physics, it is regarded as one of the most profound ideas
put forward to explain the nature of distinctive life attributes,
since it proposes, in a wide extend, the existence of very peculiar
order inherent the information and functional carriers in biology.

In earlier theoretical examination of this idea, some models of
ultrametric random walk were proposed, but they were confronted with
difficulties in applications to the protein fluctuation dynamics, in
particular, to the large body of data obtained in the experiments on
ligand-rebinding kinetics of myoglobin and spectral diffusion in
deeply frozen proteins.

Realization of this task has been provided for last years by V.
Avetisov in collaboration with A. Bikulov, S. Kozyrev, V. Osipov,
and A. Zubarev \cite{ABK,107,107a,107b,107c}. It was shown that, in
spite of extreme complexity of the protein energy landscape,
$p$-adic diffusion equation introduced in \cite{VVZ} offers a
surprisingly simple, accurate, and universal description of the
protein fluctuation dynamics on extremely large range of scales from
cryogenic up to room temperatures:
$$
{\partial f(x,t)\over \partial t}+D^{\alpha}_x f(x,t)=0,\qquad
\alpha\sim {1\over T}.
$$
Here $t$ is the real time and the $p$-adic coordinate $x$
describes the ``tree of basins'' which corresponds to the
conformational state of the protein, $T$ is the temperature. This
equation was used to describe two drastically different types of
experiments ---
 on rebinding of CO to
myoglobin and spectral diffusion of proteins.

These applications of $p$-adic diffusion equation to the protein
dynamics highlight very important protein attribute, namely, the
fact that protein dynamic states and protein energy landscape, being
both extremely complex, exhibit the hierarchical self-similarity.

On the opposite side of biological complexity, e.g. in modeling of
optimization selection of complex biological entities over
combinatorial large evolutionary spaces, the $p$-adic stochastic
processes have recently been recognized as a useful tool too. Though
the fact of natural ultrametric (taxonomic) relationships in
biology, the ultrametric representation of biological realm has not
been reflected by models.

Such an evolutionary model, based on $p$-adic diffusion equation,
has recently been proposed by V. Avetisov and Yu. Zhuravlev
\cite{107d}. It is interesting, that the model, being suggested to
describe the evolution of complex biological entities, cast light
on the basic point of the prebiotic evolutionary concepts known as
the ``error catastrophe''. It was found that the prebiotic
evolution can be getting beyond the continuity principle of
Darwinian evolutionary paradigm.

\section{{Quantization with $p$-Adic Valued Wave
Functions}}

The first step toward quantum mechanics with wave functions valued
in non-Archimedean fields (and even superalgebras) was done by
Vladimirov and Volovich \cite{VL,VL1}, see  also
\cite{Volovich:1987}. This approach was elaborated  by Khrennikov
in series of papers and books
\cite{Khrennikov:1990,Khrennikov:1991,Khrennikov:1991A,Khrennikov:1993D},
\cite{Khrennikov:1994} and extended  in collaboration with Cianci
and Albeverio
\cite{Albeverio,Albeverio2,Albeverio4,Khrennikov:1997,Khrennikov:1999}.
Here we present essentials of this theory. The basic objects of
this theory are the $p$-adic Hilbert space and symmetric operators
acting in this space \cite{Kalisch,Bayod1}. Vectors of the
$p$-adic Hilbert space which are normalized with respect to inner
product represent quantum states. $p$-Adic valued quantum theory
suffers of the absence of a ``good spectral theorem''.  At the
same time this theory is essentially simpler (mathematically),
since {\it operators of position and momentum are bounded}. Thus
the canonical commutation relations can be represented by bounded
operators\footnote{It was discovered by Albeverio and Khrennikov
\cite{Albeverio} and  later and independently by Kochubei
\cite{KU} (with a different construction of the representation),
see also Keller, Ochsenius, and Schikhof \cite{Keller}.}. It is
impossible in the complex case. Representations of groups in
Hilbert spaces are cornerstones of quantum mechanics.  In
\cite{Albeverio} Albeverio and Khrennikov constructed a
representation of the Weyl--Heisenberg group in the $p$-adic
Hilbert space. Spectra of the position and momentum
 operators were studied in \cite{Albeverio2}, \cite{Albeverio4}.

Theory of $p$-adic valued functions is exposed in the Schikhof book
\cite{schikhof1}. Spectral theory in ultrametric Banach algebras and
the ultrametric functional calculus are considered by Escassut
\cite{escassut1}, analytic functions on infraconnected sets are
studied in \cite{escassut2}. Theory of meromorphic functions over
non-Archimedean fields is presented by Pei-Chu Hu and Chung-Chun
Yang \cite{PCHu}. Differential equations for $p$-adic valued functions were studied by many
authors, see \cite{dwork,gorbachuk}.
$p$-Adic summability of perturbation series is
also investigated, see \cite{bran1,bran2,bran3}.

\section{{$\mathbb{Q}_p$-valued Probability}}

This quantum model induces $\mathbb{Q}_p$-valued
``probabilities''. Surprisingly we can proceed in the rigorous
probabilistic framework even in such unusual situation --
including limit theorems and theory of randomness, see Khrennikov
\cite{Khrennikov:1993D,Khrennikov:1994,Khrennikov:1997,Khrennikov:1999}.
The starting point of such a generalized probabilistic model was
extension of von Mises' frequency probability theory to  $p$-adic
topologies: frequencies of realizations should stabilize not with
respect to the ordinary real topology, but one of $p$-adic
topologies. We emphasize that relative frequency is always a
rational number: $\nu_n(\alpha)= n(\alpha)/N,$ where $N$ is the
total number measurements and $n$ is the number of realizations
favorable  to the fixed result $\alpha.$ In the simplest case
$\alpha=0,1.$ Thus the $p$-adic version of frequency probability
theory describes a new class of random sequences, a new type of
randomness.

\section{{Applications to Cognitive Science and Psychology}}

The idea of encoding  mental states by numbers has very  long
history -- Plato, Aristotle, Leibnitz, and the others. A new
realization of this idea was provided by Khrennikov who used rings
of $m$-adic numbers $\mathbb{Z}_m$ to encode states of the brain,
see \cite{Khrennikov:1997}. This approach was developed in
collaboration with Albeverio, Kloeden, Tirozzi, Gundlach,
Dubischar, Steinkamp
 \cite{Khrennikov2,Khrennikov3,AlbeverioC,Dubischar,Khrennikov4,Khrennikov5,Tirozzi,Khrennikov6,Khrennikov7,Khrennikov9,Khrennikov10},
 \cite{Khrennikov:1999}. The model of $m$-adic (and more general
ultrametric) {\it mental space} was  applied to cognitive science
and psychology, including modeling of coupling between unconscious
and conscious flows of information in the brain. Coupling with
neurophysiology was studied in \cite{Khrennikov9,Khrennikov10} --
the process of ``production'' of $m$-adic mental space by neuronal
trees. This model  can be used  for description of arbitrary
 flows of information on the basis of $m$-adic  {\it information space.}
The main distinguishing feature of this approach is a possibility to
encode hierarchical structures which are present in information by
using ultrametric topologies, treelike structures. Ultrametric
algorithmic information is considered by Murtagh \cite{FMurtagh}.

\section{{Applications to Image Analysis}}

Practically all images have (often hidden) hierarchical
structures. These structures can be represented by using $m$-adic
information spaces. In some cases the $m$-adic representation of
images essentially simplifies analysis, in particular, {\it image
recognition.} One can ignore details which belong to lower levels
of the $m$-adic hierarchy of an image. More effective algorithms
can be developed starting with the $m$-adic encoding of
information. This approach was developed by Benois-Pineau,
Khrennikov,  Kotovich, and Borzistaya
\cite{Benois,Khrennikov8,Khrennikov11}.

\section{{ $p$-Adic Wavelets}}


 $p$-Adic wavelet theory was initiated by Kozyrev  \cite{wavelets}.
Since in the $p$-adic case it is not possible to use for
construction of wavelet basis translations by elements of
$\mathbb{Z}$, it was proposed to use instead translations by
elements of the factor group $\mathbb{Q}_p/\mathbb{Z}_p$.
 An example of $p$-adic wavelet was introduced in the form of
the product of the character and the characteristic function
$\Omega$ of the unit ball:
$$
\psi(x)=\chi(p^{-1}x)\Omega(|x|_p).
$$
The orthonormal basis $\{\psi_{\gamma n j}\}$ of $p$-adic wavelets
in $L^2(\mathbb{Q}_p)$ was constructed \cite{wavelets} by
translations and dilations of $\psi$:
$$
\psi_{\gamma nj}(x)=p^{-{\gamma\over 2}} \chi(p^{\gamma-1}j
(x-p^{-\gamma}n)) \Omega(|p^{\gamma} x-n|_p),
$$
$$
\gamma\in \mathbb{Z},\quad n=\sum_{l=k}^{-1}n_l p^l\in
\mathbb{Q}_p/\mathbb{Z}_p,\quad n_l=0,\dots,p-1,\quad j=1,\dots,p-1.
$$

In $p$-adic analysis wavelet bases are related to the spectral
theory of pseudodifferential operators. In particular, the above $p$-adic
wavelets are eigenvectors of the Vladimirov operator:
$$
D^{\alpha}\psi_{\gamma nj}= p^{\alpha(1-\gamma)}\psi_{\gamma nj}.
$$
Spectra of more general operators were studied in \cite{nhoper}.

Moreover, the orbit of the wavelet $\psi$ with respect to the action
of the $p$-adic affine group
$$
G(a,b)f(x)=|a|_p^{-{1\over 2}} f\left({x-b\over a}\right),\qquad a,b
\in\mathbb{Q}_p,\quad a\ne 0,
$$
gives the set of products of $p$-adic wavelets from the basis
$\{\psi_{\gamma n j}\}$ and roots of one of the degree $p$. In
general \cite{frames}, the orbit of generic complex-valued locally constant
mean zero compactly supported function gives a frame of $p$-adic
wavelets.

Relation between real and $p$-adic wavelets is given by the $p$-adic
change of variable: the map
$$
\rho:\mathbb{Q}_p \to \mathbb{R}_+,\quad
\rho:\sum_{i=\gamma}^{\infty} x_i p^{i} \mapsto
\sum_{i=\gamma}^{\infty} x_i p^{-i-1},\quad x_i=0,\dots,p-1,\quad
\gamma \in \mathbb{Z}
$$
maps (for $p=2$) the basis $\{\psi_{\gamma n j}\}$ of $p$-adic
wavelets onto the basis of Haar wavelets on the positive half--line.

Important contributions in $p$-adic wavelet theory were done by
Albeverio, Benedetto, Khrennikov, Shelkovich, Skopina
\cite{wave1,wave2,wave3,wave4,wave5,wave6}. $p$-Adic Tauberian and
Shannon-Kotelnikov theorems were investigated by Khrennikov and
Shelkovich \cite{shelkovich}.


 \section{{ Analysis on General  Ultrametric
 Spaces}}


The analysis of wavelets and pseudodifferential operators on
general locally compact ultrametric spaces was developed in
\cite{General1,General2,General3,KozyrevBook}. A
pseudodifferential operator on ultrametric space $X$ is defined by
Kozyrev as the following integral operator:
$$
Tf(x)=\int_{X}T(\sup(x,y))(f(x)-f(y))\,d\nu(y)
$$
where $\nu$ is a Borel measure and some integrability condition for
the integration kernel is satisfied. The $\sup(x,y)$ is the minimal
ball in $X$ which contains the both points $x$ and $y$, i.e. the
integration kernel $T$ is a locally constant (for $x\ne y$) function
with the described domains of local constancy.

Ultrametric wavelet bases $\{\Psi_{Ij}\}$ were introduced and it was
found that ultrametric wavelets are eigenvectors of ultrametric
pseudodifferential operators:
$$
T\Psi_{Ij}=\lambda_I\Psi_{Ij}.
$$
Let us note that locally compact ultrametric spaces under
consideration are completely general and do not possess any group
structure.

Analysis on $p$-adic infinite-dimensional spaces was developed by
Kochubei, Kaneko and Yasuda  \cite{[K],Koch,[L],[M],[N]}.

Another important class of ultrametric spaces consists of locally
compact fields of positive characteristic. Analysis over such fields
has both common and different features with $p$-adic analysis. See
the book of Kochubei \cite{[O]} for the details.

\section{{ Cascade Models of Turbulence }}

One of the problems of fully developed turbulence is the
energy-cascading process. This process is characterized by large
hierarchy of scales involved and is known as the Richardson cascade.
The main idea of the Fischenko and Zelenov paper \cite{[A1]} was to
consider the equation of cascade model not in  coordinate space, but
in an ultrametric space connected with hierarchy of turbulence
eddies. This idea leads to a rather simple nonlinear integral
equation  for the velocity field.

Properties of fully developed turbulence such as multifractal
behavior of energy dissipation or  Kolmogorov's 1/3 behavior are
obtained by analysis of solutions of the nonlinear equation over
$p$-adic numbers. So, $p$-adic numbers  provide  a natural and
systematic approach to cascade models of turbulence.

A modification of nonlinear ultrametric integral equation of \cite{[A1]} was investigated in \cite{[B1]} by
Kozyrev. It was found \cite{[B1]} that, using the ultrametric wavelet analysis,
we get a family of exact solutions for this modified nonlinear equation. Moreover, for this equation an exact solution of an
arbitrary Cauchy problem with the initial condition in the space of
locally constant mean zero compactly supported functions was
constructed.

\section{{Disordered
Systems and Spin Glasses}}


Spin glasses (disordered magnetics) are typical examples of
disordered systems. Order parameter for a spin glass in the replica
symmetry breaking approach is described by the Parisi matrix ---
some special hierarchical block matrix. It was found that \cite{MPV}
the structure of correlation functions for spin glasses are related
to ultrametricity.

Avetisov, Bikulov, Kozyrev \cite{ABK} and Parisi, Sourlas
\cite{PaSu}  found that the Parisi matrix possesses the $p$-adic
parametrization, namely matrix elements of this matrix after some
natural enumeration of the lines and columns of the matrix can be
expressed as the real valued function of the $p$-adic norm of
difference of the indices
$$
Q_{ab}=q(|a-b|_p).
$$
This allows to express the correlation functions of the spin glass
in the state with broken replica symmetry in the form of some
$p$-adic integrals.

Also more general replica solutions related to general locally
compact ultrametric spaces were obtained
\cite{ReplicaI,ReplicaII,ReplicaIII}. The $p$-adic Potts model is
considered in \cite{mukhamedov1,mukhamedov2}.

         \section{{ $p$-Adic Dynamical Systems}}

The theory of  $p$-adic dynamical systems \cite{MK} is an
intensively developing discipline on the boundary between various
mathematical theories --  dynamical systems,
 number theory, algebraic geometry,
non-Archimedean analysis --  and having numerous applications --
theoretical physics, cognitive science, cryptography, computer
science,  automata theory, genetics, numerical analysis and image
analysis.

One of the sources of theory of $p$-adic dynamical systems was
dynamics in  rings of $\mod{p^n}$-residue classes.   We can
mention investigations of W. Narkiewicz, A. Batra, P. Morton and
P. Patel, J. Silverman and G. Call, D.K. Arrowsmith, F. Vivaldi
and Hatjispyros, J. Lubin, T. Pezda, H-C. Li, L. Hsia,  see, e.g.,
\cite{VIV2}, \cite{VIV1} and also books \cite{MK}, \cite{AN} for
detailed bibliography. Another flow was induced in algebraic
geometry. This (algebraic geometric) dynamical flow began with
article of M. Herman and J.C. Yoccoz \cite{Herman/Yoccoz:1981} on
the problem of small divisors in non-Archimedean fields. It seems
that this was the first publication on non-Archimedean dynamics.
In further development
 of this dynamical flow the crucial role was played by J. Silverman, R. Benedetto,
J. Rivera-Letelier,  C. Favre,  J-P. B\'ezivin, see Silverman's book
\cite{Silverman:07} for bibliography.

Another flow towards algebraic dynamics has $p$-adic theoretical
physics as its source.   One of the authors of this review  used
this pathway towards $p$-adic dynamical systems, from study of
quantum models with $\mathbb{Q}_p$-valued functions. As a result, a
research group on non-Archimedean dynamics was created at the
V\"axj\"o University, Sweeden: Andrei Khrennikov, Karl-Olof Lindahl,
Marcus Nilsson, Robert Nyqvist, and Per-Anders Svensson \cite{MK}.

We point out recent publications of V. Arnold, e.g.,  \cite{ARN5},
devoted to chaotic aspects of arithmetic dynamics closely coupled to
the problem  of turbulence.  Some adelic aspects of linear
fractional dynamical systems are considered in \cite{dragov14}.

Finally, we point out a flow towards algebraic dynamics which is
extremely important for applications to computer science,
cryptography, and numerical analysis, especially in connection with
pseudorandom numbers and uniform distribution of sequences. This
flow arose in 1992 starting with works of Anashin
\cite{me:conf,me:1}, followed by a series of his works on $p$-adic
ergodicity as well as on  above applications. It worth mention here
one of the most recent papers from these, \cite{me-spher} that
contains a solution of a problem on perturbed monomial mappings on
$p$-adic spheres -- the problem was put by A.Khrennikov, e.g.,
\cite{MK}, see  \cite{AN} for general presentation of $p$-adic
ergodic theory.

\section{{$p$-Adic Models of the Genetic Code}}


Recently, a new interesting application of $m$-adic information
space was found in the domain of genetics, $(m = 2, 4, 5$ depending
on a model); see B. Dragovich  and A. Dragovich
\cite{genetic_code0}, A. Khrennikov \cite{KHR11}, M. Pitk\"anen
\cite{Pitkanen2}, A. Khrennikov and S. Kozyrev \cite{genetic_code}.

The relation between 64 codons, which are building blocks of genes,
and 20 amino acids, which are building blocks of proteins, is known
as the genetic code, which is degenerate.  Since G. Gamow, there has
been a problem of theoretical foundation of experimentally known (
only about 16) codes in  all living organisms. The central point of
$p$-adic approach to the genetic code is identification  C=1, A=2,
T=3, G= 4,  where C, A, T, G are nucleotides in DNA  and 1, 2, 3, 4
are digits in 5-adic representation of codons, which are
trinucleotides. Using $p$-adic distances between codons, it was
shown that degeneration of  the vertebral mitochondrial code has
$p$-adic structure, and all other codes can be regarded as slight
modifications of this one. Details of this approach can be found in
\cite{dragov15} of the present volume and \cite{dragov16}.

The following $p$-adic model of the genetic (amino acid)
code was proposed in \cite{genetic_code}. It was shown that after some
$p$-adic parametrization of the space of codons the degeneracy of
the amino acid code is equivalent to the local constancy of the map
of a $p$-adic argument. In two-dimensional $2$-adic parametrization
of \cite{genetic_code} we get the following table of amino acids on
the $2$-adic plane of codons:
$$
\begin{array}{|c|c|c|c|}
\hline \begin{array}{c} {\rm Lys}  \cr \hline{\rm Asn}
\end{array}  & \begin{array}{c}
{\rm Glu}  \cr \hline{\rm Asp}
\end{array}  & \begin{array}{c}
{\rm Ter}  \cr \hline{\rm Ser}
\end{array}  & {\rm Gly}\cr
\hline\begin{array}{c} {\rm Ter}  \cr \hline{\rm Tyr}
\end{array}  & \begin{array}{c}
{\rm Gln}  \cr \hline{\rm His}
\end{array}  & \begin{array}{c} {\rm Trp}  \cr
\hline{\rm Cys}
\end{array}  & {\rm Arg}\cr
\hline \begin{array}{c} {\rm Met}  \cr \hline{\rm Ile}
\end{array}
 & {\rm Val} & {\rm Thr} & {\rm Ala}    \cr
\hline
\begin{array}{c}
{\rm Leu}  \cr \hline{\rm Phe}
\end{array}
 & {\rm Leu}  & {\rm Ser}
 & {\rm Pro} \cr \hline
\end{array}
$$

\section{{Applications to Economics, Finance, Data
Mining}}

Possible applications of $p$-adic analysis in economics, finance,
business connections, including a $p$-adic version of the
Black-Scholes equation \cite{trel}, are discussed in
\cite{khok,trel,bikul,joksim}.

Application of $p$-adic analysis in data analysis and data mining is
explored by Murtagh \cite{murtFirst}.

 \section*{{Acknowledgments}}

The authors would like to thank many colleagues for kind and
fruitful discussions of various themes of $p$-adic mathematical
physics and related topics, before and during work on this review
article.

The work of B. Dragovich  was partially supported by the Ministry
of Science and Technological Development, Serbia, under contract
No 144032D. ~ I. V. Volovich and S. V. Kozyrev gratefully
acknowledge being partially supported by the grant DFG Project 436
RUS 113/951,
 the grant   RFFI
08-01-00727-a,  the grant of the President of the Russian
Federation for the support of scientific schools NSh 3224.2008.1,
the Program of the Department of Mathematics of the Russian
Academy of Science ``Modern problems of theoretical mathematics'',
and by the program of Ministry of Education and Science of Russia
``Development of the scientific potential of High School, years of
2009--2010'', project 3341. S. V. Kozyrev also has been partially
supported by the grants DFG Project 436 RUS 113/809/0-1 and RFFI
05-01-04002-NNIO-a. Authors of this review were also partially
supported by the grants of the Swedish Royal Academy of Science
and Profile Mathematical Modeling of V\"axj\"o University.


\begin{thebibliography}{199}


\bibitem{Ser}
J.~P.~Serre,  \textit{A Course in Arithmetics} (Springer GTM 7,
1973).

\bibitem{Vol1}
I.~V.~Volovich, ``Number theory as the ultimate physical theory,''
Preprint No. TH 4781/87, CERN, Geneva, (1987).



\bibitem{Volovich:1987}
I.~V.~Volovich,  ``$p$-Adic string,'' Class. Quant. Grav.
\textbf{4}, L83--L87 (1987).


\bibitem{VL} V.~S.~Vladimirov and I.~V.~Volovich,  ``Superanalysis. I. Differential calculus,''
Theor. Math. Phys.  {\bf 59}, 317--335  (1984).

\bibitem{VL1}  V.~S.~Vladimirov and I.~V.~Volovich, ``Superanalysis. II. Integral calculus,''
Theor. Math. Phys.  {\bf 60}, 743--765 (1985).

\bibitem{Manin} Yu.~I.~Manin, ``Reflections on arithmetical
physics,'' in \textit{Conformal Invariance and String Theory}, pp.
293--303 (Academic Press, Boston, 1989).


\bibitem{Var} V.~S.~Varadarajan, ``Arithmetic Quantum Physics: Why, What, and
Whither,'' Proc.  Steklov Inst.  Math. \textbf{245}, 258--265
(2004).


\bibitem{bogol} N.~N.~Bogolyubov, ''On a new method in the theory of
superconductivity,'' J. Exp. Theor. Phys. \textbf{34} (1),
58 (1958).



\bibitem{nambu} \textit{Broken Symmetry. Selected Papers of Y. Nambu}
Eds.  T. Eguchi and K. Nishijima,
(World Scientific, Singapore, 1995).

\bibitem{VVZ}
  V.~S.~Vladimirov,   I.~V.~Volovich and  E.~I.~Zelenov, \textit{$p$-Adic
Analysis and Mathematical Physics}  (World Scientific, Singapore,
1994).

\bibitem{Khrennikov:1994}
A.~Yu.~Khrennikov,  \textit{$p$-Adic Valued Distributions in
Mathematical Physics} (Kluwer, Dordrecht, 1994).


\bibitem{BrFr}
L.~Brekke and P.~G.~O.~Freund, ``$p$-Adic numbers in physics,'' Phys.
Rep. \textbf{233} (1), 1--66 (1993).

\bibitem{Koch}  A.~N.~Kochubei, \textit{Pseudo-Differential Equations and Stochastics over Non-Archimedean
Fields}  (Marcel Dekker, New York, USA, 2001).



\bibitem{KozyrevBook} S.~V.~Kozyrev, \textit{Methods and Applications
of  Ultrametric and $p$--Adic Analysis: From Wavelet Theory to
Biophysics}, Modern Problems of Mathematics  \textbf{12} (Steklov
Math. Inst., Moscow, 2008)
http://www.mi.ras.ru/spm/pdf/012.pdf [in Russian].


\bibitem{GSW} M.~B.~Green, J.~H.~Schwarz and E.~Witten,
\textit{Superstring Theory: Volumes 1, 2} (Cambridge Univ. Press,
Cambridge, 1987).

\bibitem{FrOl}  P.~G.~O.~Freund and  M.~Olson,   ``Non-archimedean
strings,''  Phys. Lett.~B  \textbf{199},  186--190 (1987).

\bibitem{GGP}  I.~M.~Gelfand, M.~I.~Graev and I.~I.~Pyatetskii-Shapiro, \textit{Representation
Theory and Automorphic Functions} (Saunders, Philadelphia, 1969).


\bibitem{fram1} P.~H.~Frampton and Y.~Okada, ``Effective scalar field theory of $p$-adic
string,'' Phys. Rev. D \textbf{37}, 3077--3084 (1988).


\bibitem{freund} L.~Brekke, P.~G.~O.~Freund,  M.~Olson  and E.~Witten, "Nonarchimedean
string dynamics," Nucl. Phys. B \textbf{302} (3), 365--402 (1988).

\bibitem{FroWit}  P.~G.~O.~Freund and E.~Witten, ``Adelic string amplitudes,`` Phys.
 Lett. B \textbf{199}, 191--194  (1987).


\bibitem{VolHar}  I.~V.~Volovich, ``Harmonic analysis and $p$-adic strings,'' Lett.
Math. Phys. \textbf{16},  61--67 (1988).


\bibitem{borevich} Z.~I.~Borevich and I.~R.~Shafarevich, \textit{Number
Theory} (AP, 1966).


\bibitem{Vlaade}
V.~S.~Vladimirov, ``Adelic formulas for four-particle string and
superstring tree amplitudes in one-class quadratic fields,'' Proc.
Steklov Inst. Math. \textbf{245}, 3--21 (2004).


\bibitem{ADV1} I.~Ya.~Aref'eva, B.~Dragovich and I.~V.~Volovich, ``On the adelic string
amplitudes,'' Phys. Lett. B \textbf{209}, 445--450 (1988).


\bibitem{ADV2} I.~Ya.~Aref'eva, B.~Dragovich and I.~V.~Volovich,
``Open and closed $p$-adic strings and quadratic extensions of
number fields,'' Phys. Lett. B \textbf{212}, 283--289 (1988).


\bibitem{ADV3} I.~Ya.~Aref'eva, B.~Dragovich and I.~V.~Volovich,
``$p$-Adic superstrings,'' Phys. Lett. B \textbf{214}, 339--346
(1988).


\bibitem{leb}  D.~R.~Lebedev and A.~Yu.~Morozov, ``$p$-Adic single-loop
calculations,'' Theor. Math. Phys.  \textbf{82} (1), 1--6 (1990).

\bibitem{chek}  L.~Chekhov, A.~Mironov and A.~Zabrodin, ``Multiloops calculations in
$p$-adic string theory and Bruhat-Tits trees,'' Commun. Math. Phys.
\textbf{125}, 675 (1989).

\bibitem{FraOka}
P.~H.~Frampton and Y.~Okada,  ``$p$-Adic string N-point function,''
Phys. Rev. Lett.  \textbf{60}, 484-486 (1988).

\bibitem{nishino} P.~H.~Frampton and H. Nishino, ``Theory of $p$-adic
closed strings,'' Phys. Rev. Lett. \textbf{62}, 1960--1964 (1989).

\bibitem{volovich1} I.~V.~Volovich,   ``$p$-Adic space-time and string
theory,'' Theor. Math. Phys.  \textbf{71}  (3), 574--576 (1987).

\bibitem{Volmot} I.~V.~Volovich, ``From $p$-adic strings to etale ones,'' Trudy Steklov
Math. Inst. \textbf{203}, 41-47 (1994); arXiv:hep-th/9608137.

\bibitem{marcolli1} A.~Connes  and M.~Marcolli, ``Quantum fields and
motives,'' J.  Geom. Phys. \textbf{56} (1), 55--85 (2005).

\bibitem{marcolli2}    A.~Connes, C.~Consani and M.~Marcolli, ``Noncommutative
geometry and motives: the thermodynamics of endomotives,'' Advances
in Mathematics \textbf{214} (2), 761--831 (2007);
arXiv:math.QA/0512138.

\bibitem{marcolli3} A.~Connes, C.~Consani and M.~Marcolli, ``The Weil proof and the
geometry of the adeles class space,''  arXiv:math/0703392.

\bibitem{marcolli4} A.~Connes and M.~Marcolli, ``Noncommutative geometry, quantum fields,
and motives,''  Colloquium Publications  \textbf{55} (American Math.
Society, 2008).

\bibitem{marcolli5} C.~Consani and M.~Marcolli, ``Spectral triples from Mumford
curves,'' Int. Math. Research Notices \textbf{36}, 1945--1972
(2003).

 \bibitem{marcolli6} G.~Cornelissen, M.~Marcolli, K.~Reihani and A.~Vdovina, ``Noncommutative geometry on
trees and buildings,''  in \textit{Traces in Geometry, Number
Theory, and Quantum Fields}, pp. 73--98 (Vieweg Verlag, 2007).

\bibitem{voevodsky} V.~Voevodsky, ``Motives over simplicial
schemes,'' arXiv:0805.4431.


\bibitem{thakur}  M.~Baker, J.~Teitelbaum, B.~Conrad, K.~S.~Kedlaya and D.~S.~Thakur,
\textit{$p$-Adic Geometry: Lectures from the 2007 Arizona Winter
School} (American Mathematical Society, 2008).

\bibitem{[A]}  A.~ N.~Kochubei and M.~R.~Sait-Ametov, ``Interaction measures on the
space of distributions over the field of $p$-adic numbers,'' Infin.
Dimens. Anal. Quantum Probab. Related Topics \textbf{6}, 389--411
(2003).


\bibitem{Mis}
M.~D.~Missarov,  ``Random fields on the adele ring and Wilson's
renormalization group,'' Annales de l'institut Henri Poincare (A):
Physique Theorique \textbf{50} (3), 357--367 (1989).



\bibitem{Smi} V.~A.~Smirnov, ``Calculation of general $p$-adic Feynman
amplitude,'' Comm. Math. Phys. \textbf{149} (3), 623--636 (1992).



\bibitem{sen} D.~Ghoshal and A.~Sen, ``Tachyon condensation and brane
descent relations in $p$-adic string theory,'' Nucl. Phys. B
\textbf{584}, 300--312  (2000).




\bibitem{zwiebach}  N.~Moeller and B.~Zwiebach, ``Dynamics with infinitely many time
derivatives and rolling tachyons,'' JHEP \textbf{10}, 034 (2002).




\bibitem{ABGKM} I.~Ya.~Aref'eva, D.~M.~Belov, A.~A.~Giryavets, A.~S.~Koshelev and P.~B.~Medvedev,
    ``Noncommutative field theories and (super)string field
    theories,''  arXiv:hep-th/0111208.

\bibitem{ABKM} I.~Ya.~Aref'eva, D.~M.~Belov, A.~S.~Koshelev and P.~B.~Medvedev,
  ``Tachyon condensation in cubic superstring field theory,''
  Nucl. Phys. B   \textbf{638}, 3--20 (2002);
  arXiv:hep-th/0011117.

  \bibitem{Yaroslav}  Ya.~I.~Volovich, ``Numerical study of nonlinear equations
  with infinite number of derivatives,''
 J. Phys. A: Math. Gen \textbf{36}, 8685 (2003); arXiv:math-ph/0301028.

  \bibitem{VSV-Yaroslav} V.~S.~Vladimirov and Ya.~I.~Volovich,
``Nonlinear dynamics equation in $p$-adic string theory,'' Theor.
Math. Phys. \textbf{138}, 297--307 (2004); arXiv:math-ph/0306018.

\bibitem{AJK}
I.~Ya.~Aref'eva, L.~V.~Joukovskaya and A.~S.~Koshelev, ``Time
evolution in superstring field theory on non-BPS brane. Rolling
tachyon and energy-momentum conservation,'' JHEP \textbf{09}, 012 (2003);
arXiv:hep-th/0301137.

\bibitem{Vlanon}
V.~S.~Vladimirov, ``On the non-linear equation of a $p$-adic open
string for a scalar field,''   Russ. Math. Surv. \textbf{60},
1077--1092 (2005).


\bibitem{VlaFirst}   V.~S.~Vladimirov,  ``On the equations for
$p$-adic closed and open strings,''  $p$-Adic Numbers, Ultrametric
Analysis and Applications \textbf{1} (1),  79--87 (2009).



\bibitem{Trento}  V.~Forini, G.~Grignani, G.~Nardelli,  ``A new
rolling tachyon solution of cubic string field theory,''
JHEP \textbf{0503}, 079 (2005); arXiv:hep-th/0502151.


\bibitem{Proxor} D.~V.~Prokhorenko, ``On some nonlinear integral
equation in the (super)string theory,'' arXiv:math-ph/0611068.


\bibitem{Are-cosmology} I.~Ya.~Aref'eva, ``Nonlocal string tachyon as
a model for cosmological dark energy,'' AIP Conf. Proc. \textbf{826}, 301--311
(2006); arXiv:astro-ph/0410443.



\bibitem{Are-Jouk}
 I.~Ya.~Aref'eva and L.~V.~Joukovskaya, ``Time lumps in  nonlocal stringy models
and cosmological applications,'' JHEP \textbf{10}, 087 (2005);
arXiv:hep-th/0504200.

 \bibitem{AK} I.~Ya.~Aref'eva and A.~S.~Koshelev, ``Cosmic acceleration
and crossing of $w={}-1$ barrier in non-local cubic
  superstring field theory model,'' JHEP \textbf{0702}, 041 (2007);
  arXiv:hep-th/0605085.

\bibitem{Calcagni} G.~Calcagni,
``Cosmological tachyon from cubic string field theory,'' JHEP
\textbf{05},  012 (2006); arXiv:hep-th/0512259.


\bibitem{dragov13} B.~Dragovich, ``$p$-Adic and adelic quantum cosmology:
$p$-Adic origin of dark energy and dark matter,'' in
\textit{$p$-Adic Mathematical Physics},  Amer. Inst. Phys. Conf.
Series \textbf{826}, 25--42 (2006); arXiv:hep-th/0602044.


\bibitem{Cline} N.~Barnaby, T.~Biswas and J.~M.~Cline, ``$p$-Adic inflation,''
JHEP \textbf{0704}, 056 (2007);
arXiv:hep-th/0612230.

\bibitem{Koshel} A.~S.~Koshelev, Non-local SFT tachyon and cosmology,'' JHEP \textbf{0704},
029 (2007); arXiv:hep-th/0701103.

\bibitem{Jouk-PR} L.~V~Joukovskaya,  ``Dynamics in nonlocal cosmological
models derived from string field theory,''  Phys. Rev. D \textbf{76}, 105007
 (2007); arXiv:0707.1545.

\bibitem{CMN}  G.~Calcagni, M.~Montobbio and G.~Nardelli, ``Route to nonlocal cosmology,''
 Phys. Rev. D \textbf{76}, 126001 (2007).

\bibitem{Are-Jouk-Ver} I.~Ya.~Aref'eva,  L.~V.~Joukovskaya and S.~Yu.~Vernov,
''Dynamics in nonlocal linear models in the Friedmann-Robertson-Walker
metric,'' J. Phys. A: Math. Theor \textbf{41}, 304003 (2008).

\bibitem{Neil} N.~~Barnaby and  J.~M.~Cline, ``Predictions for nongaussianity
from nonlocal inflation,''
  JCAP \textbf{0806}, 030 (2008); arXiv:0802.3218.


\bibitem{Mulryne} N.~J.~Nunes, D.~J.~Mulryne, ``Non-linear non-local cosmology,''
arXiv:0810.5471.

\bibitem{Neil2} N.~Barnaby, ``Nonlocal inflation,'' arXiv:0811.0814.






\bibitem{AreVol} I.~Ya.~Aref'eva and I.~V.~Volovich,
``Quantization of the Riemann zeta-function and cosmology,'' Int. J.
Geom. Meth. Mod. Phys. \textbf{4}, 881--895  (2007);
arXiv:hep-th/0701284v2.

\bibitem{dragov10} B.~Dragovich, ``Zeta-nonlocal scalar fields,''
Theor. Math. Phys. \textbf{157} (3), 1671--1677 (2008);
arXiv:0804.4114.


\bibitem{haran}  M.~J.~Shai~Haran, \textit{The Mysteries of the Real Prime} (Oxford
University Press, USA, 2001).



\bibitem{AreVolNC}
I.~Ya.~Aref'eva and I.~V.~Volovich,  ``Quantum group particles and
non-Archimedean geometry,'' Phys. Lett. B \textbf{268}, 179-187 (1991).

\bibitem{kozy}    S.~V.~Kozyrev,
``The space of free coherent states is isomorphic to space of
distributions on $p$--adic numbers,'' Infin. Dimens. Anal.
 Quantum Prob. \textbf{1} (2), 349--355 (1998);
arXiv:q-alg/9706020.



\bibitem{marcolli3a}  A.~Connes and M.~Marcolli, ``From physics to number theory via
noncommutative geometry. Part I: Quantum statistical mechanics of
Q-lattices,'' in  \textit{Frontiers in Number Theory, Physics, and
Geometry, I} pp. 269--350 (Springer Verlag, 2006).


\bibitem{marcolli4a}     A.~Connes and M.~Marcolli,
``From physics to number theory via noncommutative geometry. Part
II: Renormalization, the Riemann-Hilbert correspondence, and motivic
Galois theory,'' in \textit{Frontiers in Number Theory, Physics, and
Geometry, II} pp. 617--713 (Springer Verlag, 2006).



\bibitem{schmidt} R. Schmidt, ``Arithmetic gravity and Yang-Mills theory: An approach to
adelic physics via algebraic spaces,'' arXiv:0809.3579v1.



\bibitem{raja1}  V.~S.~Varadarajan, \textit{Geometry of Quantum Theory} (Springer Verlag,
2007).



\bibitem{VVZ2}
V.~S.~Vladimirov and I.~V.~Volovich, ``$p$-Adic quantum mechanics,''
Comm. Math. Phys. \textbf{123}, 659--676 (1989).

\bibitem{VVZ3} V.~S.~Vladimirov, I.~V.~Volovich and E.~I.~Zelenov, ``The spectral theory in the
$p$-adic quantum mechanics,'' Izvestia Akad. Nauk SSSR, Ser. Mat.
\textbf{54}, 275--302 (1990).


\bibitem{zelen1} E.~ I.~Zelenov, ``The infinite-dimensional $p$-adic symplectic
group,'' Russian Acad. Sci. Izv. Math. \textbf{43}, 421-441 (1994).



\bibitem{harish-chandra}    Harish-Chandra, ``Harmonic analysis on reductive $p$-adic
groups,'' in \textit{Proc. of Symposia in Pure Mathematics}, Vol.
XXVI, Amer. Math. Soc. Providence, R. I., pp. 167--192 (1973).

\bibitem{weil}  A.~Weil, ``Sur certains groupes d'operateurs
unitaires,'' Acta Mathematica  \textbf{111},  143-211 (1964).

\bibitem{schneider}  P.~Schneider, ``Continuous representation theory of $p$-adic
Lie groups,'' Proc. ICM Madrid 2006, Vol. II, pp. 1261 - 1282
(2006).





\bibitem{zelenFirst} E.~I.~Zelenov, ``Quantum approximation theorem,''
 $p$-Adic Numbers,
Ultrametric Analysis and Applications \textbf{1} (1),  88--90
(2009).

\bibitem{varadFirst} V.~S.~Varadarajan,  ``Multipliers for the symmetry groups
of $p$-adic spacetime,''     $p$-Adic Numbers, Ultrametric Analysis
and Applications \textbf{1} (1), 69--78 (2009).



\bibitem{zelmas} E.~I.~Zelenov,  ``$p$-Adic Heisenberg group and Maslov index,''
Commun. Math. Phys. \textbf{155}, 489--502 (1993).


\bibitem{VV1} V.~S.~Vladimirov  and I.~V.~Volovich,
``A $p$-adic Schr\"odinger-type equation,'' Lett. Math. Phys.
\textbf{18}, 43--53 (1989).

\bibitem{[B]} A.~N.~Kochubei, ``A Schr\"odinger type equation over the field of $p$-adic
numbers,'' J. Math. Phys. \textbf{34}, 3420--3428 (1993).


\bibitem{[F]} W.~A.~Zuniga-Galindo, ``Decay of solutions of wave-type
pseudo-differential equations over $p$-adic fields,''  Publ. Res.
Inst. Math. Sci. \textbf{42}, 461--479 (2006); \textbf{44}, 45--48
(2008).


\bibitem{digernes}  T.~Digernes, V.~S.~Varadarajan and D.~Weisbart, ``Matrix valued
Schr\"odinger operators on local fields,'' to be published in Proc.
Steklov Math. Inst.  \textbf{265}, (2009).



\bibitem{dragov1} B.~Dragovich, ``Adelic model of harmonic
oscillator,'' Theor. Math. Phys. \textbf{101}, 1404--1415 (1994);
arXiv:hep-th/0402193.

\bibitem{dragov2} B.~Dragovich, ``Adelic harmonic oscillator,'' Int.
J. Mod. Phys. A \textbf{10}, 2349--2365 (1995);
arXiv:hep-th/0404160.

\bibitem{dragov3} B.~Dragovich,   ``$p$-Adic and adelic quantum mechanics,''
Proc.  Steklov Inst. Math. \textbf{245}, 72--85 (2004);
arXiv:hep-th/0312046v1.

\bibitem{dragov4} G.~Djordjevi\'c, B.~Dragovich and Lj.~Ne\v si\'c,
``$p$-Adic and adelic free relativistic particle,'' Mod. Phys. Lett.
A \textbf{14}, 317--325 (1999); arXiv:hep-th/0005216 .

\bibitem{dragov17} B.~Dragovich, ``On generalized functions in adelic quantum
mechanics,''  Integral Transform. Spec. Funct. \textbf{6}, 197-203
(1998); arXiv:math-ph/0404076.


\bibitem{radyno} E.~M.~Radyna and Ya.~V.~Radyno,
``Distributions and mnemofunctions on adeles,'' Proc. Steklov. Inst.
Math. \textbf{245},   215-227 (2004).


\bibitem{Parisi} G.~Parisi, ``$p$-Adic functional integral,'' Mod. Phys. Lett. A \textbf{4},
369--374 (1988).

\bibitem{zelen2}  E.~I.~Zelenov, ``$p$-Adic path integrals,'' J. Math. Phys. \textbf{32}, 147--153 (1991).

\bibitem{varadarajan}  V.~S.~Varadarajan, ``Path integrals for a class of $p$-adic Schr\"odinger
equations,'' Lett. Math. Phys. \textbf{39}, 97--106  (1997).

\bibitem{smolyanov} O.~G.~Smolyanov and N.~N.~Shamarov,  ``Feynman and
Feynman-Kac formulas for evolution equations with Vladimirov
operator,''  Doklady Mathematics  \textbf{77}  (3), 345-350 (2008).


\bibitem{dragov7} B.~Dragovich, ``Adelic wave function of the
Universe,'' in \textit{Proc. Third A. Friedmann Int. Seminar on
Grav. and Cosmology}, Eds.  Yu.~N.~Gnedin, A.~A.~Grib and
V.~M.~Mostepanenko, pp. 311--321 (Friedmann Lab. Publishing, St.
Petersburg, 1995).


\bibitem{dragov8} G.~Djordjevi\'c, B.~Dragovich and Lj.~Ne\v si\'c,
``Adelic path intergals for quadratic Lagrangians,'' Infin. Dimens.
Anal. Quan. Prob. Relat. Topics \textbf{6}, 179--195 (2003);
arXiv:hep-th/0105030.

\bibitem{dragov9}  G.~Djordjevi\'c and B.~Dragovich,
``$p$-Adic path integrals for quadratic actions,'' Mod. Phys. Lett.
A \textbf{12} (20), 1455--1463 (1997); arXiv:math-ph/0005026.

\bibitem{fernandez}   R.~ N.~Fernandez, V.~S.~Varadarajan and D. Weisbart,
``Airy functions over local fields,'' to be publ. in Lett. Math.
Phys.



\bibitem{ADFV}
 I.~Ya.~Aref'eva, B.~Dragovich, P.~H.~Frampton and I.~V.~Volovich,
``The wave function of the Universe and $p$-adic gravity,'' Int. J.
Mod. Phys. A \textbf{6}, 4341--4358  (1991).


\bibitem{dragov5} B.~Dragovich, ``Adelic generalization of wave function of
the Universe,'' The First Hungarian-Yugoslav Astronomical
Conference, Hungary, Baja, April 26-27, 1995. Publ. Obs. Astron.
Belgrade \textbf{49} 143-144 (1995).

\bibitem{DDNV}  G.~S.~Djordjevi\'c, B.~Dragovich, Lj.~D.~Ne\v si\'c
and I.~V.~Volovich, ``$p$-Adic and adelic minisuperspace quantum
cosmology,''  Int. J. Mod. Phys. A \textbf{17} (10), 1413--1433
(2002); arXiv:gr-qc/0105050.

\bibitem{dragov6} B.~Dragovich and Lj. Ne\v si\'c, ``$p$-Adic and adelic generalization of
quantum cosmology,'' Gravitation and Cosmology \textbf{5}, 222-228
(1999); arXiv:gr-qc/0005103.








\bibitem{[G]}   R.~S.~Ismagilov, ``On the spectrum of the self-adjoint operator in
$L_2(K)$ where $K$ is a local field; an analog of the Feynman-Kac
formula,'' Theor. Math. Phys. \textbf{89}, 1024--1028 (1991).

\bibitem{[H]}  A.~N.~Kochubei, ``Parabolic equations over the field of $p$-adic
numbers,'' Math. USSR Izv. \textbf{39}, 1263--1280 (1992).

\bibitem{[I]} S.~Haran, ``Analytic potential theory over $p$-adics,'' Ann. Inst.
Fourier \textbf{43}, 905-944 (1993).


\bibitem{BikVol} A.~Kh.~Bikulov and I.~V.~Volovich, ``$p$-Adic Brownian
motion,'' Izv. Math. \textbf{61} (3), 537--552 (1997).


\bibitem{randomfield} A.~Yu.~Khrennikov and S.~V.~Kozyrev, ``Ultrametric random
field,'' Infin. Dimens. Anal. Quan. Prob. Related Topics \textbf{9}
(2), 199--213 (2006); arXiv:math/0603584.


\bibitem{kamizono} K.~Kamizono, ``$p$-Adic Brownian motion over
$\mathbb{Q}_p$,'' to be published.

\bibitem{evans} S.~N.~Evans, ``Local field
Brownian motion,'' J. Theor. Probab. \textbf{6}, 817--850 (1993).


\bibitem{karwowski}  S.~Albeverio and W.~Karwowski,
``A random walk on $p$-adic numbers - generator and its spectrum,''
Stochastic processes. Theory and Appl. \textbf{53}, 1 - 22 (1994).



\bibitem{yasuda1} K.~Yasuda, ``Additive processes on local fields,''
 J. Math. Sci. Univ.
Tokyo  \textbf{3}, 629-654 (1996).



\bibitem{belopolskaya}  S.~Albeverio and Ya.~Belopolskaya,
``Stochastic processes in $\mathbb{Q}_p$ associated with  systems of
nonlinear PDEs,'' $p$-Adic Numbers, Ultrametric Analysis and
Applications, to be published (2009).





\bibitem{[C]} H.~Kaneko and A.~N.~Kochubei, ``Weak solutions of stochastic
differential equations over the field of $p$-adic numbers,'' Tohoku
Math. J. \textbf{59}, 547--564 (2007).



\bibitem{KanFirst} H.~Kaneko, ``Fractal theoretic aspects of local field,''    $p$-Adic Numbers,
Ultrametric Analysis and Applications \textbf{1} (1),  51--57
(2009).


\bibitem{vladUMN}   V.~S.~Vladimirov, ``Generalized functions over the field of $p$-adic
numbers,''
 Russ. Math. Surv. \textbf{43},  19-64  (1988).


\bibitem{[D]}   A.~N.~Kochubei, ``A non-Archimedean wave equation,'' Pacif. J. Math.
\textbf{235}, 245--261 (2008).

\bibitem{[E]}  W.~A.~Zuniga-Galindo, ``Parabolic equations and Markov processes
over $p$-adic fields,'' Potential Anal. \textbf{28}, 185--200
(2008).

\bibitem{torba} S.~Albeverio, S.~Kuzhel and S.~Torba, ``$p$-Adic
Schr\"odinger-type operator with
point interactions,'' J. Math. Anal. Appl. \textbf{338}, 1267-1281 (2008).

\bibitem{RTV} G.~ Rammal, M.~A.~ Toulouse and M.~A.~ Virasoro, ''Ultrametricity for physicists,''
Rev. Mod. Phys. \textbf{58}, 765-788 (1986).

\bibitem{ABK}  V.~A.~Avetisov, A.~H.~Bikulov and S.~V.~Kozyrev,
``Application of $p$--adic analysis to models of spontaneous
breaking of  replica symmetry,''  J. Phys. A: Math. Gen. \textbf{32}
(50), 8785--8791  (1999); arXiv:cond-mat/9904360 .

\bibitem{107}  V.~A.~Avetisov, A.~Kh.~Bikulov, S.~V.~Kozyrev and
V.~A.~Osipov,    ``$p$--Adic models of ultrametric diffusion
constrained by hierarchical energy landscapes,'' J. Phys. A: Math.
Gen. \textbf{35} (2), 177--189   (2002); arXiv:cond-mat/0106506 .



\bibitem{107a} V.~A.~Avetisov, A.~Kh.~Bikulov and V.~A.~Osipov, `` $p$--Adic
models for ultrametric diffusion in conformational dynamics of
macromolecules,'' Proc. Steklov Inst. Math. \textbf{245}, 48--57
(2004).

\bibitem{107b} V.~A.~Avetisov and A.~Kh.~Bikulov, ``Protein ultrametricity and
spectral diffusion in deeply frozen proteins,'' in press, Biophys.
Rev. and Lett. \textbf{3}  (3),  (2008); arXiv:0804.4551.

\bibitem{107c} V.~A.~Avetisov, A.~Kh.~Bikulov and A.~P.~Zubarev, ``First passage
time distribution and number of returns for ultrametric random
walk,'' in press, J. Phys. A: Math. Theor. \textbf{42} (2009);
arXiv:0808.3066.

\bibitem{107d} V.~A.~Avetisov and Yu.~N.~Zhuravlev, ``An evolutionary
interpretation of the $p$--adic ultrametric diffusion equation,''
Doklady Mathematics \textbf{75} (3), 435-455 (2007); arXiv:0808.3066.



\bibitem{Khrennikov:1990}
A.~Yu.~Khrennikov,  ``Mathematical methods of the
non-ar\-chi\-me\-de\-an physics,''  Uspekhi Mat. Nauk \textbf{45}
(4),  79--110 1990.

\bibitem{Khrennikov:1991}
A.~Yu.~Khrennikov,   ``$p$-Adic quantum mechanics with $p$-adic
valued functions,''  J. Math. Phys. \textbf{32}, 932--937 (1991).

 \bibitem{Khrennikov:1991A}
A.~Yu.~Khrennikov,  ``Real-non-Archimedean structure of
space-time,'' Theor.  Math. Phys.  \textbf{86} (2), 177-190 (1991).

\bibitem{Khrennikov:1993D}
A.~Yu.~Khrennikov,  ``$p$-Adic probability theory and its
applications. The principle of statistical stabization of
frequencies,''   Theor. Math. Phys.  \textbf{97} (3), 348-363
(1993).


\bibitem{Albeverio}
 S.~Albeverio and  A.~Yu.~Khrennikov,
 ``Representation of the Weyl group in spaces of square integrable
functions with respect to $p$-adic valued Gaussian distributions,''
 J. Phys. A: Math.  Gen. \textbf{29}, 5515-5527 (1996).


\bibitem{Albeverio2}
S.~Albeverio, R.~Cianci and A.~Yu.~Khrennikov,  ``On the spectrum of
the $p$-adic position operator,''   J. Phys. A:  Math. Gen.
\textbf{30}, 881-889 (1997).

\bibitem{Albeverio4}
S.~Albeverio, R.~Cianci and A.~Yu.~Khrennikov, ``On the Fourier
transform and the spectral properties of the $p$-adic momentum and
Schrodinger operators,''  J. Phys. A: Math.  Gen. \textbf{30},
5767-5784 (1997).


\bibitem{schikhof1} W.~H.~Schikhof, \textit{Ultrametric Calculus} (Cambridge University Press,
Cambridge, 1984).

\bibitem{escassut1}  A.~Escassut, \textit{Ultrametric Banach Algebras} (World Scientific, Singapore,
2003).

\bibitem{escassut2} A.~Escassut, \textit{Analytic Elements in p-Adic Analysis} (World
Scientific, Singapore 1995).

\bibitem{PCHu}  P.-C. Hu and C.-C. Yang, \textit{Meromorphic Functions over
non-Archimedean Fields} (Kluwer Academic Publishers, 2001).

\bibitem{dwork} B.~Dwork, G.~Gerotto and F.~J.~Sullivan,  \textit{An
Introduction to G-Functions}
(Princeton University Press, 1994).

\bibitem{gorbachuk} M.~L.~Gorbachuk and  V.~I.~Gorbachuk,
 ``On the Cauchy problem for differential
equations in a Banach space over the field of $p$-adic numbers, I, II,'' Meth.
Funct. Anal. Topology \textbf{9}, 207--212 (2003); Proc.
Steklov Inst. Math. \textbf{245}, 91-97
(2004).


\bibitem{bran1} I.~Ya.~Aref'eva, B.~Dragovich and I.~V.~Volovich,
``On the $p$-adic summability of the anharmonic oscillator,'' Phys.
Lett. B \textbf{200}, 512--514 (1988).


\bibitem{bran2}  B. Dragovich,
``$p$-Adic perturbation series and adelic summability,'' Phys. Lett.
B \textbf{256}, 392--396 (1991).

\bibitem{bran3} B. Dragovich,
``On some $p$-adic series with factorials,'' in \textit{p-Adic
Functional Analysis}, Proc. Fourth Int. Conf.
 $p$-Adic Analysis, Eds. W.H.Schikhof et al., Lecture Notes on Pure and
 Appl. Math. \textbf{192}, pp. 95--105  ( Marcel Dekker, N.Y., 1997).

\bibitem{Khrennikov:1997}
A.~Yu.~Khrennikov,   \textit{Non-Archimedean Analysis: Quantum
Paradoxes, Dynamical Systems  and Biological Models}  (Kluwer,
Dordrecht, 1997).


\bibitem{Khrennikov:1999}
A.~Yu.~Khrennikov,  \textit{Interpretations of Probability} ( VSP,
Utrecht, 1999).

\bibitem{Kalisch}   G.~K.~Kalisch, ``On $p$-adic Hilbert spaces,'' Ann.
 Math. \textbf{48}, 180--192  (1947).

\bibitem{Bayod1} S.~Albeverio, J.~M.~Bayod, C.~Perez-Garcia,
A.~Yu.~Khrennikov and R.~Cianci, ``Non-Archimedean analogues of
orthogonal and symmetric operators,'' Izv. Akad. Nauk \textbf{63}
(6), 3-28 (1999).


\bibitem{KU} A.~N.~Kochubei,  ``$p$-Adic commutation relations,'' J. Phys.
A: Math. Gen.  \textbf{29}, 6375-6378 (1996).

\bibitem{Keller} H.~Keller, H.~Ochsenius and W.~H.~Schikhof, ``On the commutation
relation $AB-BA=I$ for operators on nonclassical Hilbert spaces,''
in  \textit{$p$-Adic Functional Analysis}, Eds. A.~K. Katseras, W.~H.~ Schikhof
and L.~van Hamme. Lecture Notes in Pure and Appl. Math.
\textbf{222}, 177-190 (2003).

\bibitem{Khrennikov2}
A.~Yu.~Khrennikov,  ``Human subconscious as the $p$-adic dynamical
system,'' J. Theor. Biology \textbf{ 193}, 179--196 (1998).

\bibitem{Khrennikov3}
A.~Yu.~Khrennikov,   ``$p$-Adic dynamical systems: description of
concurrent struggle in biological population with limited growth,''
Dokl. Akad. Nauk \textbf{361}, 752 (1998).

\bibitem{AlbeverioC}
S.~Albeverio,   A.~Yu.~Khrennikov and  P.~Kloeden,  ``Memory
retrieval as a $p$-adic dynamical system,'' Biosystems \textbf{ 49},
105--115 (1999).

\bibitem{Dubischar}
D.~Dubischar, V.~M.~Gundlach,  O.~Steinkamp and  A.~Yu.~Khrennikov,
 ``A $p$-adic model for the process of thinking disturbed by
physiological and information noise,'' J. Theor. Biology \textbf{
197}, 451--467 (1999).

\bibitem{Khrennikov4}
A.~Yu.~Khrennikov,  ``Description of the operation of the human
subconscious by means of $p$-adic dynamical systems,'' Dokl. Akad.
Nauk \textbf{365}, 458--460 (1999).

\bibitem{Khrennikov5}
A.~Yu.~Khrennikov,  ``$p$-Adic discrete dynamical systems and
collective behaviour of information states in cognitive models,''
Discrete Dynamics in Nature and Society \textbf{5}, 59--69 (2000).

\bibitem{Tirozzi} S.~Albeverio, A.~Yu.~Khrennikov and B.~Tirozzi, ``$p$-Adic neural
networks,''  Math. Models and Meth. in Appl. Sciences
\textbf{9} (9), 1417-1437 (1999).

\bibitem{Khrennikov6}
A.~Yu.~Khrennikov,   ``Classical and quantum mechanics on $p$-adic
trees of ideas,'' BioSystems \textbf{56}, 95--120 (2000).

\bibitem{Khrennikov7}
A.~Yu.~Khrennikov, \textit{Classical and Quantum Mental Models and
Freud's Theory of Unconscious Mind}, Series Math. Modelling in
Phys., Engineering and Cognitive Sciences \textbf{1} (V\"axj\"o
Univ. Press, V\"axj\"o, 2002).

\bibitem{Khrennikov9}
A.~Yu.~Khrennikov,  \textit{Information Dynamics in Cognitive,
Psychological, Social,  and Anomalous Phenomena} (Kluwer, Dordreht,
 2004).

\bibitem{Khrennikov10}
A.~Yu.~Khrennikov,  ``Probabilistic pathway representation of
cognitive information,'' J. Theor. Biology \textbf{231}, 597--613
(2004).

\bibitem{FMurtagh} F.~Murtagh, ``On ultrametric algorithmic
information,'' Computer Journal, in press. (Online, Advance Access,
9 Oct. 2007, http://dx.doi.org/10.1093/comjnl/bxm084).


\bibitem{Benois}
J.~Benois-Pineau,  A.~Yu.~Khrennikov,   and N.~V.~Kotovich,
``Segmentation of images in $p$-adic and Euclidean metrics,''
Doklady Mathematics \textbf{64} (3), 450--455 (2001).


\bibitem{Khrennikov8}
A.~Yu.~Khrennikov  and N.~V.~Kotovich,  ``Representation  and
compression of images with the aid of the $m$-adic coordinate
system,'' Dokl. Akad. Nauk \textbf{387} (2), 159--163 (2002).



\bibitem{Khrennikov11}
A.~Yu.~Khrennikov,   N.~V.~Kotovich and E.~L.~Borzistaya,
``Compression of images with the aid of representation by $p$-adic
maps and approximation by Mahler's polynomials,''  Doklady
Mathematics \textbf{69} (3), 373--377 (2004).




\bibitem{wavelets}  S.~V.~Kozyrev,  ``Wavelet theory as $p$-adic spectral
analysis,'' Izvestiya: Mathematics \textbf{66} (2), 367--376 (2002);
arXiv:math-ph/0012019 .

\bibitem{nhoper} S.~V.~Kozyrev, ``$p$-Adic pseudodifferential operators and $p$-adic
wavelets,''  Theor. Math. Phys. \textbf{138} (3), 322--332 (2004);
arXiv:math-ph/0303045 .

\bibitem{frames} S.~Albeverio and S.~V.~Kozyrev, ``Frames of $p$-adic wavelets and orbits of the affine
group,'' $p$-Adic Numbers, Ultrametric Analysis and Applications \textbf{1} (1), 18-33 (2009);  arxiv:0801.4713.

\bibitem{wave1} J.~J.~Benedetto and R.~L.~Benedetto, ``A wavelet theory for local
fields and related groups,'' J. Geom. Analysis \textbf{3}, 423--456
(2004).

\bibitem{wave2} S.~Albeverio, S.~Evdokimov and M.~Skopina, ``$p$-Adic multiresolution
analysis and wavelet frames,'' (2008);  arXiv:0802.1079v1 .

\bibitem{wave3} S.~Albeverio, S.~Evdokimov and M.~Skopina, ``$p$-Adic multiresolution analyses,''
arXiv:0810.1147 .

\bibitem{wave4} A.~Yu.~Khrennikov and V.~M.~Shelkovich, ``Non-Haar $p$-adic wavelets and
their application to pseudodifferential operators and equations,''
(2006); arXiv:0808.3338v1 .


\bibitem{wave5} A.~Yu.~Khrennikov, V.~M.~Shelkovich and M.~Skopina, ``$p$-Adic refinable
functions and MRA-based wavelets,''  J. Appr. Theory
(2008); arXiv:0711.2820 .

\bibitem{wave6} V.~M.~Shelkovich and M.~Skopina, ``$p$-Adic Haar multiresolution analysis
and pseudo-differential operators,''  J. Fourier Anal.
 Appl. (2008); arXiv:0705.2294 .


\bibitem{shelkovich} A.~Yu.~Khrennikov and V.~M.~Shelkovich, ``Distributional
asymptotics and $p$-adic Tauberian and Shannon-Kotelnikov
theorems,'' Asymptotical Analysis \textbf{46} (2), 163--187 (2006).

\bibitem{General1} S.~V.~Kozyrev and A.~Yu.~Khrennikov, ``Pseudodifferential
operators on ultrametric spaces and ultrametric wavelets,''
Izvestiya: Mathematics  \textbf{69} (5),  989-1003  (2005).

\bibitem{General2} A.~Yu.~Khrennikov and S.~V.~Kozyrev, ``Wavelets on ultrametric
spaces,'' Appl. Comp. Harmonic Analysis \textbf{19},
61--76 (2005).

\bibitem{General3} S.~V.~Kozyrev, ``Wavelets and spectral analysis of ultrametric
pseudodifferential operators,''  Sbornik: Mathematics \textbf{198}
(1), 97--116 (2007); arXiv:math-ph/0412082.


\bibitem{[K]}  A.~N.~Kochubei, ``Analysis and probability over infinite extensions of
a local field,'' Potential Anal. \textbf{10}, 305--325 (1999).

\bibitem{[L]}  K.~Yasuda, ``Extension of measures to infinite dimensional
spaces over $p$-adic field,'' Osaka J. Math. \textbf{37}, 967--985
(2000).

\bibitem{[M]}   A.~N.~Kochubei, ``Hausdorff measure for a stable-like process
over an infinite extension of a local field,''  J. Theor. Probab.
\textbf{15}, 951--972 (2002).

\bibitem{[N]}   H.~Kaneko and K.~Yasuda, ``Capacities associated with
Dirichlet space on an infinite extension of a local field,'' Forum
Math. \textbf{17}, 1011--1032 (2005).

\bibitem{[O]} A.~N.~Kochubei, \textit{Analysis in Positive Characteristic}
(Cambridge University Press, Cambridge, 2009).

\bibitem{[A1]} S.~Fischenko and E.~I.~Zelenov, ``$p$--Adic models of
turbulence,'' in \textit{$p$-Adic Mathematical Physics}, Eds.
A.~Yu.~Khrennikov, Z.~Raki\'c and I.~V.~Volovich, AIP Conference
Proceedings \textbf{286} pp. 174--191 (Melville, New York, 2006).

\bibitem{[B1]} S.~V.~Kozyrev, ``Toward an ultrametric theory of
turbulence,'' Theor.  Math. Phys.  \textbf{157} (3), 1711--1720
(2008); arXiv:0803.2719.




\bibitem{MPV}   M.~Mezard,  G.~Parisi and  M.~Virasoro,
\textit{Spin-Glass Theory and Beyond}  (World Scientific, Singapore,
1987).


\bibitem{PaSu}    G.~Parisi and  N.~Sourlas, ``$p$--Adic numbers and replica symmetry
breaking,'' European Phys. J. B \textbf{14}, 535--542 (2000);
arXiv:cond-mat/9906095 .

\bibitem{ReplicaI} A.~Yu.~Khrennikov and S.~V.~Kozyrev, ``Replica symmetry breaking related to a general ultrametric space I:
Replica matrices and functionals,'' Physica A: Stat. Mech. Appl.
\textbf{359}, 222-240  (2006); arXiv:cond-mat/0603685.

\bibitem{ReplicaII} A.~Yu.~Khrennikov and S.~V.~Kozyrev, ``Replica symmetry breaking related to a general ultrametric space II:
RSB solutions and the $n\to 0$ limit,''  Physica A: Stat. Mech.
Appl.  \textbf{359}, 241-266  (2006); arXiv:cond-mat/0603687.

\bibitem{ReplicaIII} A.~Yu.~Khrennikov and S.~V.~Kozyrev, ``Replica symmetry breaking related to a general ultrametric space
III: The case of general measure,'' Physica A: Stat. Mech. Appl.
 \textbf{378} (2), 283-298  (2007);  arXiv:cond-mat/0603694 .

\bibitem{mukhamedov1} F.~Mukhamedov and U.~Rozikov,  ``On inhomogeneous $p$-adic
Potts model on a Cayley tree,'' Infin. Dimens. Anal. Quantum Probab.
Relat. Top. \textbf{8} (2), 277--290 (2005).

\bibitem{mukhamedov2} A.~Yu.~Khrennikov, F.~M.~Mukhamedov and J.~F.~Mendes, ``On $p$-adic
Gibbs measures of the countable state Potts model on the Cayley
tree,'' Nonlinearity \textbf{20},  2923--2937 (2007).





\bibitem{MK} A.~Yu.~Khrennikov and M.~Nilsson, \textit{$p$-Adic Deterministic and Random
Dynamical Systems} (Kluwer, Dordreht, 2004).



\bibitem{VIV1} D.~K.~Arrowsmith and F.~Vivaldi, ``Geometry of $p$-adic Siegel
discs,'' Physica D \textbf{71}, 222--236 (1994).

\bibitem{VIV2} F.~Vivaldi and S.~Hatjyspyros, ``Galois theory of periodic orbits of
polynomial maps,'' Nonlinearity D \textbf{5}, 961--978 (1992).

\bibitem{AN} V.~Anashin and A.~Khrennikov,
\textit{Applied Algebraic Dynamics}, De Gruyter Expositions in
Mathematics (Walter De Gruyter Inc, Berlin, 2009).

\bibitem{Herman/Yoccoz:1981} M.~R.~Herman and J.~C.~Yoccoz, ``Generalization
of some theorem of small divisors to
 non-archimedean fields,'' in \textit{Geometric Dynamics},
 Lecture Notes Math. \textbf{1007}, pp. 408--447
(Springer-Verlag, New York--Berlin--Heidelberg, 1983).


\bibitem{Silverman:07} J.~Silverman, \textit{The Arithmetic of Dynamical
Systems}, Graduate Texts in Mathematics \textbf{241}
(Springer-Verlag, New York, 2007).




\bibitem{ARN5} V.~I.~Arnold, ``Number-theoretic
turbulence in Fermat-Euler arithmetics and large Young diagrams
geometry statistics,'' J. Math. Fluid Mech. \textbf{7}, 4-50 (2005).

\bibitem{dragov14} B.~Dragovich, A.~Khrennikov and D. Mihajlovi\'c,
``Linear fractional $p$-adic and adelic dynamical systems,'' Rep.
Math. Phys. \textbf{60}, 55--68 (2007); arXiv:math-ph/0612058.


\bibitem{me:conf} V.~S.~Anashin, ``Uniformly distributed sequences
over $p$-adic integers,'' in
  \textit{ Number Theoretic and Algebraic Methods in Computer Science},
  Eds. I.~Shparlinsky A.~J. van~der Poorten
and H.~G. Zimmer, Proc. Int. Conf. (Moscow, June--July, 1993), pp.
1--18 (World  Scientific, Singapore 1995).


\bibitem{me:1}
V.~S.~Anashin, ``Uniformly distributed sequences of $p$-adic
integers,''  Mathematical Notes \textbf{55} (2), 109--133 (1994).

\bibitem{me-spher}
V.~Anashin,  ``Ergodic transformations in the space of $p$-adic
integers,''  in  \textit{ $p$-Adic Mathematical Physics}, Eds.
A.~Yu.~Khrennikov, Z. Raki\'c and I.~V. Volovich,  AIP Conf. Proc.
\textbf{826}, pp. 3--24 (Melville, New York, 2006).




\bibitem{genetic_code0}  B.~Dragovich and A.~Yu.~Dragovich,
``A $p$-adic model of DNA sequence and genetic code,''
arXiv:q-bio.GN/0607018.

\bibitem{KHR11}  A.~Yu.~Khrennikov, ``$p$-Adic information space and gene expression,'' in \textit{Integrative
Approaches to Brain Complexity}, Eds. S.~Grant, N.~Heintz and
J.~Noebels, p. 14 (Wellcome Trust Publ., 2006).

\bibitem{Pitkanen2}
M.~Pitk\"anen,  ``Could genetic code be understood number
theoretically?'' Electronic preprint:
www.helsinki.fi/~matpitka/pdfpool/genenumber.pdf (2006).

\bibitem{dragov15} B.~Dragovich and A.~Yu.~Dragovich, ``A $p$-adic
model of DNA sequence and genetic code,'' $p$-Adic Numbers,
Ultrametric Analysis and Applications \textbf{1} (1),  34--41
(2009); arXiv:q-bio.GN/0607018 .

\bibitem{dragov16} B.~Dragovich and A.~Yu.~Dragovich, ``$p$-Adic
modelling of the genome and the genetic code,'' Computer Journal,
doi:10.1093/comjnl/bxm083, to appear (2009); arXiv:0707.3043.

\bibitem{genetic_code} A.~Yu.~Khrennikov, S.~V.~Kozyrev, ``Genetic code on the dyadic
plane,''   Physica A: Stat. Mech. Appl. \textbf{381}, 265--272
(2007); arXiv:q-bio.QM/0701007.


\bibitem{khok} M.~N.~Khokhlova and I.~V.~Volovich, ``Modeling theory and
hypergraph of classes,'' Proc.  Steklov Inst.
Math. \textbf{245}, 266--272  (2004).

\bibitem{trel} J.~Q.~Trelewicz and I.~V.~Volovich,  ``Analysis of business
connections utilizing theory of topology of random graphs,'' in
\textit{$p$-Adic Mathematical Physics}, Eds. A.~Yu.~Khrennikov,
Z.~Raki\'c and I.~V.~Volovich. AIP Conf. Proc. \textbf{826}, pp.
330-344 (Melville, New York, 2006).

\bibitem{bikul}   A.~Kh.~Bikulov, A.~P.~Zubarev and L.~V.~Kaidalova, ``Hierarchical dynamical
model of financial market near the crash point and $p$-adic
analysis,'' Vestnik Samarskogo Gosudarstvennogo Tekhnicheskogo
Universiteta. Seriya "Fiziko-Matematicheskie Nauki"  \textbf{42},
135--141 (2006) [in Russian].

\bibitem{joksim} B.~Dragovich and D.~Joksimovi\'c,  ``On possible uses
of $p$-adic analysis  in econometrics,'' Megatrend Revija \textbf{4}
(2), 5 - 16 (2007).



\bibitem{murtFirst} F.~Murtagh,  ``From data to the $p$-adic or ultrametric model,''
$p$-Adic Numbers, Ultrametric Analysis and Applications \textbf{1}
(1),  58--68 (2009).




\end{thebibliography}
\end{document}